\documentstyle[aps,prl,multicol,epsf]{revtex}
\newcommand{\ul}{\underline}
\newcommand{\op}[1]{\mbox{$\hat{#1}$}}

\begin{document}

\title{Monte Carlo simulation with time step quantification in terms
  of Langevin dynamics}

\author{U.~Nowak$^1$, R.~W.~Chantrell$^2$, and E.~C.~Kennedy$^3$}

\address{ $^1$Theoretische Tieftemperaturphysik,
  Gerhard-Mercator-Universit\"at-Duisburg, 47048 Duisburg, Germany\\ 
  $^2$ SEES, University of Wales Bangor, Dean Street, Bangor, LL57
  1UT, UK\\
  $^3$ Dept.~of Appl.~Maths and Theo.~Phys., The Queen's University of
  Belfast, Belfast, BT7 1NN, Northern Ireland\\ }
\maketitle
\begin{abstract}
  For the description of thermally activated dynamics in systems of
  classical magnetic moments numerical methods are desirable.  We
  consider a simple model for isolated magnetic particles in a uniform
  field with an oblique angle to the easy axis of the particles. For
  this model, a comparison of the Monte Carlo method with Langevin
  dynamics yields new insight in the interpretation of the Monte Carlo
  process, leading to the implementation of a new algorithm where the
  Monte Carlo step is time-quantified. The numeric results for the
  characteristic time of the magnetisation reversal are in excellent
  agreement with asymptotic solutions which itself are in agreement
  with the exact numerical results obtained from the Fokker-Planck
  equation for the N\'{e}el-Brown model.
\end{abstract}
\begin{multicols}{2}
\narrowtext

Studies of spin dynamics in particulate systems are currently of
significant interest, as model systems for understanding the
thermodynamics of the reversal process. Brown \cite{brown} developed
a theoretical formalism for thermally activated magnetisation
reversal based on the Fokker-Planck (FP) equation which led to a high
energy barrier asymptotic formula in the axially symmetric case of a
particle with easy (uniaxial) anisotropy axis collinear with the
applied magnetic field. Since then extensive calculations 
\cite{klik,braun,coffeyPRB,coffeyPRL,coffeyJP} have been carried out
in which improved approximations were found for the axially symmetric
case. Coffey and co-workers \cite{coffeyPRB,coffeyPRL,coffeyJP} also
derived formulae for the non-axially symmetric case, investigating
also the different regimes imposed by the damping parameter $\alpha$ of the
Landau-Lifshitz-Gilbert (LLG) equation. This work represents an
important basis for the understanding of dynamic processes in
single-domain particles. New experimental techniques which allow for
an investigation of nanometer-sized, isolated, magnetic particles
confirmed this theoretical approach to thermal activation
\cite{wernsdorfer}.  

Unfortunately, the extension of this work to the important case of
strongly coupled spin systems such as are found in micromagnetic
calculations of magnetisation reversal is non-trivial, and realistic
calculations in systems with many degrees of freedom would appear to
be impossible except by computational approaches.  These are currently
of two types: (i) calculations involving the direct simulation of the
stochastic (Langevin) equation of the problem, in this case the LLG
equation supplemented by a random force representing the thermal
perturbations. This is referred to as the Langevin Dynamics (LD)
formalism \cite{chant,garcia}, and (ii) Monte Carlo (MC) simulations
\cite{binder} with a continuously variable (Heisenberg like)
Hamiltonian \cite{gonzales,hinzke}.  The LD approach, although having
a firm physical basis is limited to timescales of the order of a few
ns for strongly coupled systems. The MC approach is capable of
studying longer timescales involving reversal over large energy
barriers, but has the severe problem of having no physical time
associated with each step, resulting in unquantified dynamic
behavior.

Physically, the dynamic behavior of interacting spin systems is a
topic of considerable current interest, much of this interest being
driven by the need to understand spin electronic devices such as MRAM.
The possibility of truly dynamic models of strongly coupled systems
would seem to be an important factor in the development of a
fundamental physical understanding. This requires dynamic studies over
the whole time range from ns and sub - ns to the so-called 'slow
dynamic' behavior arising from thermally excited decay of metastable
states over timescales from 10-100s and upwards.  It is inconceivable
that the LD technique can be used over the whole timescale and
therefore a truly time quantified MC technique is necessary in order
to allow calculations over the longer timescales of physical
interest. Here we propose a technique for the quantification of the MC
timestep and give a supporting argument developed from the fluctuation
dissipation theorem. This argument results in a theoretical expression
for the timestep in terms of the size of MC move, and also gives the
validity criterion that the MC timestep is much longer than the
precession time. Comparison with an analytical formula for relaxation
in the intermediate to high damping limit is used to verify the
theoretically predicted relationship relating the timestep to the size
of MC move. This represents an important first step in the process of
deriving a theoretical formalism for time quantified MC calculations
of strongly interacting spin systems.

We consider an ensemble of isolated single-domain particles where
each particle is represented by a magnetic moment with energy
\begin{equation}
  E(\ul{S}) = - d V S_z^2 - \mu_s \ul{B} \cdot \ul{S},
\end{equation}
where $\ul{S} = \ul{\mu}/\mu_s$ is the magnetic moment of unit length,
$\ul{B} = B_x \ul{\op x} + B_z \ul{\op z}$ represents a magnetic field
under an arbitrary angle $\psi$ to the easy axis of the system, $d$ is
the uniaxial anisotropy energy density and $V$ the volume of the
particle.  Throughout the article we use the material parameters $V =
8 \times 10^{-24} \mbox{m}^3$, $d = 4.2 \times 10^5 \mbox{J/m}^3$,
magnetic moment $\mu_s = 1.12 \times 10^{-17} \mbox{J/T}$.

The LLG equation of motion with LD \cite{brown} is
\begin{equation}
  \ul{\dot{S}} = - \frac{\gamma}{(1 + \alpha^2)
    \mu_s} \ul{S} \times \Big( \ul{H}(t) + \alpha \ul{S} \times
    \ul{H}(t) \Big),
\end{equation}
where $\gamma = 1.76 \cdot 10^{11} (Ts)^{-1}$ is the gyromagnetic ratio,
$\ul H(t) = \ul{\zeta}(t)-\frac{\partial E}{\partial \ul{S}}$, 
and $\zeta$ is the thermal noise with $\langle \zeta_i(t)
\rangle = 0$ and $\langle \zeta_i(t) \zeta_j(t') \rangle = \delta_{i
  j} \delta(t-t') 2 \alpha k_B T \mu_s / \gamma$.  $i$ and $j$ denote
the cartesian components.

The equation above is solved numerically using the Heun method
\cite{garcia}.  Also, it is possible to obtain analytically asymptotic
solutions for the escape rate which have been extensively compared
with the exact numerical solutions from the corresponding matrix form
of the FP equation for a wide range of parameters and non-axially
symmetric potentials \cite{coffeyPRB,coffeyPRL,coffeyJP}.

Both of our simulations, MC as well as LD, start with the magnetic
moments in $z$-direction. The magnetic field has a negative
$z$-component so that the magnetization will reverse after some time.
The time that is needed for the $z$-component of the magnetization to
change its sign averaged over a large number of runs ($N=1000$) is the
characteristic time $\tau$ which corresponds to the inverse of the
escape rate following from exact numerical solutions of the
corresponding FP equation.

For the MC simulations we use a heat-bath algorithm. The
trial step of our MC algorithm is a random movement of the
magnetic moment within a cone with a given size.  In order to achieve
this efficiently we construct a random vector with constant
probability distribution within a sphere of radius $R$. This random
vector is added to the initial moment and subsequently the resulting
vector is normalized.

\begin{figure}
  \epsfxsize=65mm\epsffile{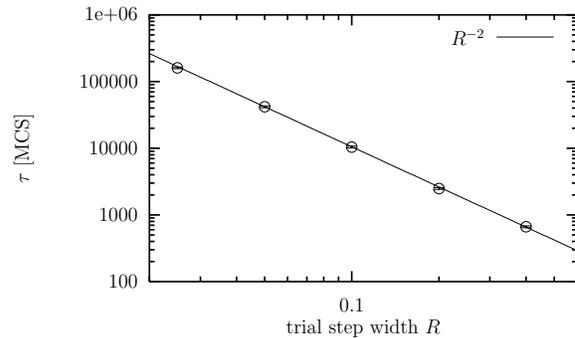}
  \caption{Characteristic time versus trial step width for a MC
    simulation. The solid line is fitted, yielding $\tau \sim
    R^{-2}$.} 
  \label{f:tofr}
\end{figure}

The size of the cone $R$ of our algorithm influences the time scale
the method simulates.  We investigate the influence of $R$ on our MC
algorithm by varying $R$ and calculating $\tau$. As usual in a MC
procedure the time is measured in Monte Carlo steps (MCS). For our
calculation we use a field of $|\ul{B}| = 0.2$T and an angle of $\psi
= 27^{\circ}$ to the easy axis. The resulting energy barrier is
$\Delta E = 8.2 \times 10^{-19}$J, the temperature we chose for
Fig.~\ref{f:tofr} is $\Delta E/k_B T = 3.3$.  As Fig. \ref{f:tofr}
demonstrates, it is $\tau \sim R^{-2}$.  This dependence can be
understood by considering the moments as performing a random walk
where $R$ is proportional to the mean step width. Having understood
that the MC time can be set by choosing an appropriate size of the
step width we search for a relation for $R$ such that one MCS
corresponds to a real-time interval, in the sense of LD.

MC methods calculate trajectories in phase space following a
master equation which describes the coupling of a system to the heat
bath.  Hence, only the irreversible part of the dynamics of the system
is considered \cite{reif} --- there is no precession of the moments
since no equation of motion is solved during the simulation.
Nevertheless, in the following we will argue that the exact knowledge
of the movement of the single moments is not necessary in order to
describe the effects of thermal activation in an ensemble of systems
under the following conditions: (i) the relevant time scales are
larger than the precession time $t_p$ of the moments, (ii) we consider
the high damping limit of the LLG equation where the energy
dissipation during one cycle of the precession is considerably large
so that the system relaxes (to the local energy minimum) on the same
time scale $t_r \approx t_p$.
\begin{figure}
  \hspace*{5mm} \epsfxsize=65mm\epsffile{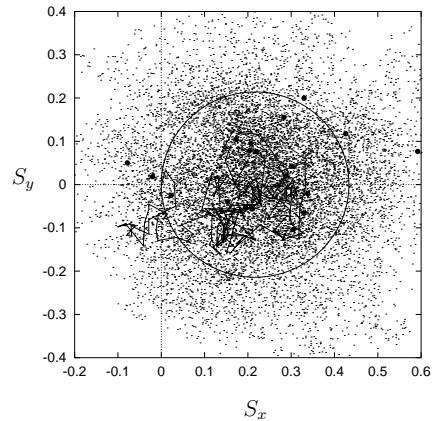}\\
  \caption{Configuration in phase space $(S_x,S_y)$ of an ensemble of 20
    particles following from a LD simulation for $\alpha = 1$: 
    see text for details.}
  \label{f:ph}
\end{figure}

\begin{figure}
  \epsfxsize=75mm\epsffile{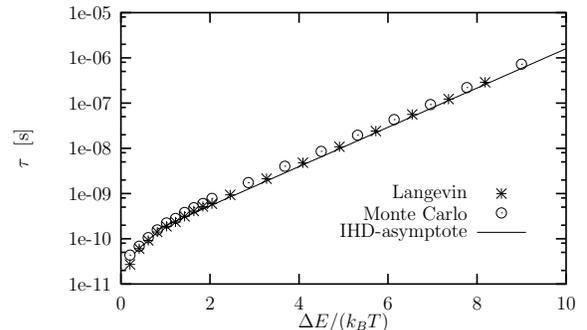}
  \caption{$\tau$ versus inverse temperature: comparison of IHD
    asymptote, LD simulation, and MC simulation.}
  \label{f:toft-a1}
\end{figure}

In Fig.~\ref{f:ph} we present the time evolution of our system in
phase space, $(S_x,S_y)$, following from a simulation of the LLG
equation for high damping, $\alpha = 1$. We use $\Delta E/k_B T =
8.2$, a rather low temperature so that the characteristic time $\tau$
for the escape from the local energy minimum is of the order of
$10^{-6}$s (see also Fig.~\ref{f:toft-a1}). The spin-precession time
is $t_p = 9 \times 10^{-11}$s here. The simulation starts close to
the local energy minimum with $S_x = S_y = 0, S_z = 1$ and the solid
line shows the trajectory of one moment over a time interval of
$\Delta t = t_p$. The 20 points are the positions of an ensemble of
20 moments after the same time.  As one can see, the moments show no
significant precession (the precession of an undisturbed moment,
i.~e.~without relaxation and fluctuations is indicated by the circle
around the energy minimum at $S_y = 0, S_x \approx 0.22$).  The small
dots represent 1000 states of the ensemble for $t < 6 \times t_p$.
Altogether, Fig.~\ref{f:ph} demonstrates that in the high damping case
already after time periods of only a few $t_p$ the moments are
uncorrelated and the ensemble reaches a local equilibrium
configuration (remember that the time scale to leave the local
equilibrium is much larger here so that Fig.\ref{f:ph} shows only the
local short-time equilibration, not the escape from the local energy
minimum).

We will show that this high-damping scenario can also be simulated by
a MC simulation and we will now derive a relation for $R$ in
order to quantify the MC time step. The intention is to
compare the fluctuations which are established in the MC
technique within one MCS with the fluctuations within a given time
scale associated with the linearized LLG equation.  Close to a local
energy minimum one can write the energy, given that first order terms
vanish as
\begin{equation}
  E \approx E_0 + \frac{1}{2}\sum_{i,j} A_{i j} S_i S_j,
  \label{e:energy}
\end{equation}
where the $S_i$ are the variables representing small deviations from
equilibrium.  In our system, for $B_x = 0$ we find equilibrium along
the $z$ axis, leading to variables $S_x$ and $S_y$. The energy
increase $\Delta E$ associated with fluctuation in $S_x$ and $S_y$ is
$\Delta E \approx \frac{1}{2} ( A_{xx} S_x^2 + A_{yy} S_y^2)$, with
$A_{xx} = A_{yy} = 2 d V + \mu_s B_z$.  Rewriting the LLG equation in
the linearized form, $\dot{S}_x = L_{xx}S_x + L_{xy}S_y$,
$\dot{S}_y = L_{yx}S_x + L_{yy}S_y$, it has been shown
\cite{lyberatos} that the correlation function takes the form
\begin{equation}
  \label{e:mu}
  \langle S_{i}(t) S_{j}(t') \rangle = \mu_{i j}
  \delta_{i, j} \delta(t-t').
\end{equation}
Dirac's $\delta$ function is here an approximation for exponentially
decaying correlations on time scales $t-t'$ that are much larger than
the time scale of the exponential decay $t_r$.  The covarianz matrix
$\mu_{i j}$ can be calculated from the system matrices $\mathbf A$ and
$\mathbf L$ as \cite{lyberatos} $ \mu_{i j} = -k_B T ( L_{i k} A_{k
  j}^{-1} + L_{j k} A_{k i}^{-1})$.  For our problem a short
calculation yields $\mu_{xx} = \mu_{yy} = 2 k_B T \frac{\alpha
  \gamma}{(1+\alpha^2) \mu_s}$.  Integrating the fluctuating
magnetisation $S_x(t)$ over a finite time interval $\Delta t$, Eq.
\ref{e:mu} takes the form
\begin{equation}
  \langle \overline{S}_x^2 \rangle = \mu_{xx} \Delta t = 2 k_B T \frac{\alpha
  \gamma}{(1+\alpha^2) \mu_s} \Delta t ,
\end{equation}
representing the fluctuations of $S_x$ averaged over a time interval
$\Delta t$.

Next, we calculate the fluctuations $\langle S_x^2 \rangle$
during one MCS of a MC simulation. This is possible if we
assume that all magnetic moments are initially in their equilibrium
position.  For our MC algorithm described above the
probability distribution for trial steps with step width $r =
\sqrt{S_x^2 + S_y^2}$ is $p_{\mbox{\tiny t}} = 3 \sqrt{R^2-r^2}/(2
\pi R^3)$. The acceptance probability within a heat bath algorithm is
$p_{\mbox{\tiny a}}(r) = 1/(1+\exp(\Delta E(r^2) / k_B T))$,
where $\Delta E(r^2)$ can be taken from Eq. \ref{e:energy}.  Hence, for
the fluctuations within one MC step it is:
\begin{equation} 
\langle S_x^2 \rangle = 2 \pi \int_0^R r \; \mbox{d} r \; \frac{r^2}{2} \;
       p_{\mbox{\tiny t}}(r) p_{\mbox{\tiny a}}(r) =
       \frac{R^2}{10} + {\cal O}(R^4)
\end{equation}
where the last line is an expansion for small $R$. By equalizing the
fluctuations within corresponding time intervals we find the relation
\begin{equation}
  R^2 = \frac{20 k_B T \alpha \gamma}{(1+\alpha^2) \mu_s} \Delta t.
  \label{e:trial}
\end{equation}
Note, from our derivation above it follows that one time step $\Delta
t$ must be larger than the intrinsic time scale $t_r$ of the
relaxation. This means - as already mentioned above - that the Monte
Carlo method can only work on time scales that are much larger than
any microscopic time scale of a precession or relaxation (to local
equilibrium) of the moment.

In principle, equation \ref{e:trial} gives the possibility to choose
the trial step for a MC simulation in such a way that 1MCS
corresponds to a real time interval, say $\Delta t = 10^{-12}$s.
However, there are of course restrictions for possible values of $R$,
like $R < 1$.  Also, $R$ should not be too small since then a Monte
Carlo algorithm is inefficient.  Therefore, either one has to choose
such a value for $\Delta t$ so that $R$ takes on reasonable values
(these will usually be of the order of $10^{-12}$s) or one uses a
reasonable constant value for $R$, say 0.1, and uses Eq.
\ref{e:trial} to calculate $\Delta t$ as the real time interval
corresponding to 1MCS.  In the following we use the first method since
it turns out to be very efficient to change $R$ with temperature.
However, we confirmed that the other method yields the same results.

To test the validity of our considerations we performed MC
simulations with an algorithm using a trial step according to
Eq.~\ref{e:trial} with $\Delta t \approx 6 \times 10^{-12}$s (the
inverse value of $\gamma$, in other words the time in the LLG equation
is rescaled by $\gamma$).  For Fig. \ref{f:toft-a1} we set $\alpha =
1$ and compare the data for $\tau(T)$ following from our MC
simulation with results from LD simulations and with
the intermediate to high damping (IHD) asymptote
\cite{coffeyPRL,coffeyJP}, namely
\begin{equation}
\tau = \frac{2 \pi \omega_{0}}{\Omega_{0} \omega_{2}} e^{\beta
  (V_0-V_2)} = \frac{2 \pi \omega_{0}}{\Omega_{0} \omega_{2}}
  e^{\Delta E / k_{B} T}, 
\end{equation}
where $\omega_0$ and $\Omega_0$ are the saddle and damped saddle
angular frequencies which have been defined in Eqs. (21) and (22) of
Ref. \cite{coffeyJP} explicitly. $\omega_2$ is the well angular
frequency for the deeper of the two potential wells and is defined in
Eq. (20) of Ref.  \cite{coffeyJP}. All have been defined in terms of
the coefficients of the truncated Taylor series representation of the
energy equation described in detail in section V of Ref.
\cite{coffeyPRB}, (particularly Eqs. 59-64). For the purpose of
comparison with MC and LD simulations, we consider one escape path
only, $e^{\beta (V_0-V_2)}$, where $\beta = V / k_BT$ and $V_0-V_2$
is the energy described by Eq.  (62) of Ref.  \cite{coffeyPRB}. For
our purposes, $\beta (V_0-V_2)$ may be represented by $\Delta E / k_B
T$.  The validity condition for the IHD formula is $\alpha \Delta E /
k_B T \gg 1$ where $\Delta E / k_B T > 1$ which have been satisfied in
all cases represented here.

From Fig. \ref{f:toft-a1} it is clear that the LD data agree with the
asymptote above. For higher temperatures the asymptote is no longer
appropriate. Here, the numerical data for $\tau$ tend to zero for $T
\to \infty$ as one expects. The MC data deviate slightly and are
roughly 10\% larger. However, considering the fact that to the best of
our knowledge this is the first comparison of a "real-time MC
simulation" with LD simulations and asymptotic formulae, the agreement
is remarkable - especially taking into account the simple form of
Eq.\ref{e:trial} underlying our algorithm and also that there is no
adjusted parameter in all our calculations and formulae.
\begin{figure}
  \epsfxsize=75mm\epsffile{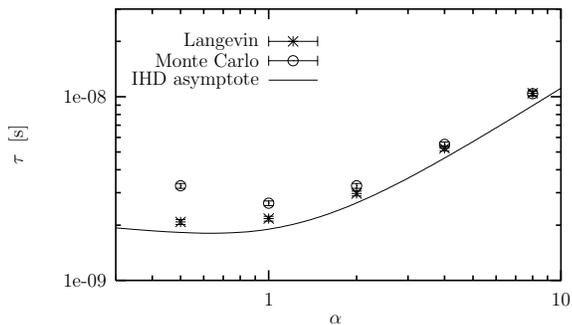}
  \caption{$\tau$ versus damping constant: comparison of IHD
    asymptote, LD simulation, and MC simulation.}
  \label{f:tofa}
\end{figure}

Since we expect that our MC procedure leads to a high damping
limit we also tested the $\alpha$-dependence of $\tau$.
Fig.~\ref{f:tofa} shows the corresponding data for the same parameter
values as before and $\Delta E / T = 3.3$. The figure shows that the
MC data converge to the IHD formula and to the data from
LD simulation for large $\alpha$. Even the small 10\%
deviation of the MC data mentioned before (Fig.
\ref{f:toft-a1}) vanishes in the limit of larger $\alpha$.

To summarize, we discussed the conditions under which a comparison of
LD with a MC process appears to be possible.  Considering a simple
system of isolated single-domain particles we derived an equation for
the trial step width of the MC process so that one step of the MC
algorithm can be related to a certain time interval. Testing this
algorithm we found excellent agreement with data from LD simulation as
well as with intermediate to high damping asymptotes for the
characteristic times of the magnetisation reversal.  Even, though our
algorithm was derived only for the special system which we consider
here, we belive that the arguments we brought forward might the
fundament even for the MC simulation of more complicated systems,
especially systems consisting of interacting magnetic moments.

\acknowledgments We would like to thank W.\ T.\ Coffey and K.\ D.\ 
Usadel for helpful discussions.  E.\ C.\ Kennedy thanks EPSRC for
financial support (GR/L06225).  R.\ W.\ Chantrell thanks EPSRC for
financial support (ref GR/M24011). This work was done within the
framework of the COST action P3 working group 4.

\end{multicols}

\end{document}